# Fixed Rank Kriging for Cellular Coverage Analysis

Hajer Braham, Sana Ben Jemaa, Gersende Fort, Eric Moulines and Berna Sayrac

***Abstract*—Coverage planning and optimization is one of the most crucial tasks for a radio network operator. Efficient coverage optimization requires accurate coverage estimation. This estimation relies on geo-located field measurements which are gathered today during highly expensive drive tests (DT); and will be reported in the near future by users' mobile devices thanks to the 3GPP Minimizing Drive Tests (MDT) feature [1]. This feature consists in an automatic reporting of the radio measurements associated with the geographic location of the user's mobile device. Such a solution is still costly in terms of battery consumption and signaling overhead. Therefore, predicting the coverage on a location where no measurements are available remains a key and challenging task. This paper describes a powerful tool that gives an accurate coverage prediction on the whole area of interest: it builds a coverage map by spatially interpolating geo-located measurements using the Kriging technique. The paper focuses on the reduction of the computational complexity of the Kriging algorithm by applying Fixed Rank Kriging (FRK). The performance evaluation of the FRK algorithm both on simulated measurements and real field measurements shows a good trade-off between prediction efficiency and computational complexity. In order to go a step further towards the operational application of the proposed algorithm, a multicellular use-case is studied. Simulation results show a good performance in terms of coverage prediction and detection of the best serving cell.**

***Keywords*—Wireless Network, Coverage Map, Radio Environment Map, Spatial Statistics, Fixed Rank Kriging, Expectation-Maximization algorithm.***

## I. Introduction

Coverage planning and optimization is one of the most crucial tasks for a radio network operator. Efficient coverage optimization requires accurate coverage estimation. This estimation relies on geo-located field measurements, gathered today during highly expensive drive tests (DT) and will be reported in the near future by users' mobile devices thanks to the 3GPP Minimization of Drive Tests (MDT) feature standardized since Release 9 [2]. The radio measurements together with the best possible geo-location will be then automatically reported to the network by the user's mobile device. Thanks to the integration of Global Positioning System (GPS) in the new generation of users' mobile devices, the geo-location information is quite accurate. Hence, with MDT, the network operator will soon have at his disposal a rich source of information that provides a greater insight into the end-user perceived quality of service and a better knowledge of the radio environment.

H. Braham is with Orange Labs research center, Issy-Les-Moulineaux, France and Télécom ParisTech, Paris, France.

S. Ben Jemaa and B. Sayrac are with Orange Labs research center, Issy-Les-Moulineaux, France.

G. Fort and E. Moulines are with LTCI Télécom ParisTech & CNRS, Paris, France.

The collection and exploitation of location aware radio measurements was introduced much earlier in the literature in the context of the cognitive radio paradigm [3]. The radio Environmental Map (REM) concept was introduced by Zhao [4] as a database that stores geo-located radio environmental information mainly for opportunistic spectrum access. The REM concept was then extended to an entity that not only stores geo-located radio information but also post processes this information in order to build a complete map. The missing information, namely the considered radio metric in locations where no measurements are available, is then predicted by interpolating the geo-located measurements [5]–[7].

The REM was then studied in the framework of European Telecommunications Standards Institute (ETSI) as a tool for the exploitation of geo-located radio measurements for the radio resource management of mobile wireless networks. A technical report dedicated to the definition of use-cases for building and exploiting the REM gives the following definition [8]: "*The Radio Environment Map (REM) defines a set of network entities and associated protocols that trigger, perform, store and process geo-located radio measurements (received signal strength, interference levels, Quality of Service (QoS) measurements [...]) and network performance indicators. Such measurements are typically performed by user equipments, network entities or dedicated sensors.*" In this ETSI report, several use-cases for REM exploitation in radio resource management are described such as coverage and capacity optimization, and interference management especially for the introduction of a new technology.

Inspired by the geo-statistics area, Kriging technique was applied to REM construction, mainly for coverage prediction and analysis in radio mobile networks [9]–[11]. Bayesian Kriging was first applied to 3G Received Signal Code Power (RSCP) coverage prediction in [9], then to Long Term Evolution (LTE) Reference Signal Received Power (RSRP) coverage analysis in [10]. The description of the bayesian Kriging methodology and the algorithm used in [9], [10], is detailed in [11]. These papers give promising results in terms of performance. However the computational complexity of the algorithm increases cubically with the number of measurement points ($\sim O(N^3)$, where $N$ is the number of measurement points).

In this paper, we aim at providing a method for predicting LTE RSRP coverage map based on MDT data. Given the huge number of measurements that will be reported by mobile terminals with MDT in the near future, reducing the computational complexity of the REM construction becomes crucial. In [12], [13], we used the Fixed Rank Kriging (FRK) introduced by Cressie in [14] (also called in the literature Spatial Random Effects model), as a method to reduce the computational complexity of the Kriging technique applied to radio coverage prediction; the method was evaluated on simulated data (see [12]) and on real field data (see [13]), both in the situation of a single cell with an omni-directional antenna. In this paper, we go a step further towards operational application of the REM prediction algorithm by considering a multicellular use-case: the directivity of the antennas is introduced in the model, and both the coverage prediction and the good detection of the best serving cell are part of the statistical analysis.

The contribution of this paper can be summarized in the following:

- We describe the FRK algorithm and its adaptation to radio coverage data. It requires an estimation step of the unknown parameters of the model: we show that the method of moments proposed in [14] can not apply and we derive a Maximum Likelihood alternative.
- We extend our model to a multicellular use-case with directive antennas.
- We evaluate the performances of the proposed algorithms both on simulated and real data.

The paper is organized as follows: Section II starts with an overview of the propagation models existing in the literature. Then the statistical parametric model is introduced. The last part is devoted to the parameter estimation: the applicability of the original method is discussed, and an alternative is given. In Section III, the extension to the multicellular use-case is detailed. Then the numerical analysis in the single cell and multicellular use-cases are provided in Section IV. Finally, Section V summarizes the main conclusions.

## II. Radio Environment Map prediction models

In this section, we give an overview of basic propagation models and give some notations that will be used in the remainder of this paper. Then we introduce a new model for REM construction, which is adapted from the FRK model proposed in [14].

### A. Introduction to propagation modeling and notations

A radio propagation model describes a relation between the signal strength, and the locations of the transmitter and the receiver. There are in the literature two different approaches for this description which are respectively derived using analytical and empirical methods [15]. The analytical approach is based on fundamental principals of the radio propagation concept. The empirical one introduces a statistical model and uses a set of observations to fit this model. The advantage of the second approach is the use of actual field measurements to estimate the parameters of the model.

Denote by $Z(x)$ the received power at the receiver end located at $x \in \mathbb{R}^2$, expressed in dB. The *path-loss model*, also called in the literature *the log-distance model*, is among the analytical approaches. It describes $Z(x)$ as a logarithmically decreasing function of the distance $\mathrm{dist}(x)$ between the transmitter location and the receiver location $x$ (see e.g. [15]):

$$Z(x) = p_t - 10\kappa \ln_{10}(\mathrm{dist}(x)), \qquad x \in \mathbb{R}^2; \qquad (1)$$

$p_t$ is the transmitted power in dB and $\kappa$ is the path loss exponent. When using this formula to predict the REM, $p_t$ is considered as known since it is one of the antenna characteristic, and $\kappa$ depends on the propagation environment. For example, $\kappa$ is in the order of 2 in free space propagation and it is larger when considering an environment with obstacles (see e.g. [15], [16]).

The model in Eq. (1) does not take into account the fact that two mobile Equipment (ME) equally distant from the base station (BS), may have different environment characteristics. To tackle this bottleneck, empirical approaches based on a statistical modeling of the shadowing effect have been introduced. The *log-normal shadowing model* consists in setting (see [17])

$$Z(x) = p_t - 10\kappa \ln_{10}(\mathrm{dist}(x)) + \sigma_\nu \tilde{\nu}(x), \qquad x \in \mathbb{R}^2, \quad (2)$$

where $(\tilde{\nu}(x))_x$, introduced to capture the shadowing effect, is a standard Gaussian variable (note that the terminology "log-normal" comes from the fact that the shadowing term expressed in dB is normally distributed), and $\sigma_\nu > 0$. With this model, the REM prediction at location $x$ is $\hat{Z}(x) = p_t - 10\kappa \ln_{10}(\mathrm{dist}(x))$. The unknown parameters $p_t$ and $\kappa$ are estimated from measured data, usually by the maximum likelihood estimator (which is also the least-square estimator in this Gaussian case).

Both the models (1) and (2) are large-scale propagation models: they do not consider the small fluctuations of the received power due to the local environment. The *correlated shadowing model* captures these small-scale variations:

$$Z(x) = p_t - 10\kappa \ln_{10}(\mathrm{dist}(x)) + \nu(x), \qquad x \in \mathbb{R}^2, \quad (3)$$

where $(\nu(x))_x$ is a zero mean Gaussian process with a parametric spatial covariance function $(C(x,x'))_{x,x'}$. This model implies that two signals $Z(x), Z(x')$ at different locations $x, x'$ are correlated, with covariance equal to $C(x,x')$. The REM prediction formula based on the model (3) is known in the literature as the Kriging (see e.g. [18]): the prediction $\hat{Z}(x)$ is the conditional expectation of $Z(x)$ given the measurements. It depends linearly on these measurements (see [18, Eq. (3.2.12)]) and involves a computational cost $O(N^3)$, where $N$ is the number of measurement points. Here again, the prediction necessitates the estimation of the parameters: different parameter estimation approaches were proposed (see e.g. [18], [19] for maximum likelihood, or [11], [18] for a Bayesian approach). This model was applied to REM interpolation in [11], [19], [20] and this technique has proved to realize accurate prediction performances.

All the models above assume that the antennas are omni-directional. Nevertheless, in macro-cellular networks, operators usually deploy directional antennas. Hence, the received power depends also on the direction of reception. To fit the model to this new constraint, several papers proposed to modify the model (2) by adding a term $\bar{G}(x)$ depending on the mobile location $x$ and modeling the antenna gain (see e.g. [21], [22]): for $x \in \mathbb{R}^2$,

$$Z(x) = p_t - 10\kappa \ln_{10}(\mathrm{dist}(x)) + \bar{G}(x) + \nu(x). \qquad (4)$$

Different gain functions $\bar{G}$ are proposed, depending on the antenna used for the transmission (for example, a polar antenna, a sectorial antenna, ...); see e.g. [22]–[24]. The function $\bar{G}$ depends on parameters which are usually considered known; we will allow the function $\bar{G}$ to depend on unknown parameters to be calibrated from the observations. In this paper, we will extend the model (4) by considering a correlated spatial noise $\nu(x)$.

### B. Fixed Rank Kriging prediction model

For $x \in \mathbb{R}^2$, $Z(x)$ is assumed of the form

$$Z(x) = p_t - 10\kappa \ln_{10} \mathrm{dist}(x) + \varsigma G(x) + \boldsymbol{s}(x)^T \boldsymbol{\eta}, \quad (5)$$

where $\boldsymbol{s} : \mathbb{R}^2 \to \mathbb{R}^r$ collects $r$ deterministic spatial basis functions and $\boldsymbol{\eta}$ is a $\mathbb{R}^r$-valued zero mean Gaussian vector with covariance matrix $\boldsymbol{K}$. $A^T$ denotes the transpose of the matrix $A$ and by convention, the vectors are column-vectors. $p_t - 10\kappa \ln_{10} \mathrm{dist}(x) + \varsigma G(x)$ describes the large scale spatial variation (i.e. the trend) and the random process $(\boldsymbol{s}(x)^T \boldsymbol{\eta})_x$ is a smooth small-scale spatial variation. In practice, the number of basis functions $r$ and the basis functions $\boldsymbol{s}$ are chosen by the user (see [14, Section 4] and Section IV-B1 below). It is assumed that the function $G$ is known: in the case of an omni-directional antenna, $G$ is the null function, and for directional antenna we give an example in Section III.

We have $N$ measurement points $y_1, \cdots, y_N$ modeled as the realization of the observation vector $\mathbf{Y} = (Y(x_1), \ldots, Y(x_N))^T$ at known locations $x_1, \cdots, x_N$ and defined as follows

$$Y(x) = Z(x) + \sigma\, \varepsilon(x), \qquad x \in \mathbb{R}^2. \quad (6)$$

$(\varepsilon(x))_x$ is assumed to be a zero mean standard Gaussian process, it incorporates the uncertainties of the measurement technique. $\boldsymbol{\eta}$ and $(\varepsilon(x))_x$ are assumed to be independent so that the covariance matrix of $\mathbf{Y}$ is given by

$$\boldsymbol{\Sigma} = \sigma^2 \boldsymbol{I}_N + \boldsymbol{S}\boldsymbol{K}\boldsymbol{S}^T, \quad (7)$$

where $\boldsymbol{S} = (\boldsymbol{s}(x_1), \ldots, \boldsymbol{s}(x_N))^T$ is the $N \times r$ matrix, and $\boldsymbol{I}_N$ denotes the $N \times N$ identity matrix. This model implies that the conditional distribution of $(Z(x))_x$ given the observations $\mathbf{Y}$ is a Gaussian process. Its expectation and covariance functions are respectively given by (see e.g. [25, Appendix A.2])

$$x \mapsto t^T(x)\boldsymbol{\alpha} + \boldsymbol{s}(x)^T \boldsymbol{K}\boldsymbol{S}^T \boldsymbol{\Sigma}^{-1}(\mathbf{Y} - \boldsymbol{T}\boldsymbol{\alpha}), \quad (8)$$

$$(x, x') \mapsto \boldsymbol{s}^T(x)\boldsymbol{K}\boldsymbol{s}(x') - \boldsymbol{s}(x)^T \boldsymbol{K}\boldsymbol{S}^T \boldsymbol{\Sigma}^{-1} \boldsymbol{S}\boldsymbol{K}\boldsymbol{s}(x'), \quad (9)$$

$$\text{where} \quad \boldsymbol{T} = \begin{bmatrix} 1 & -10 \ln_{10} \mathrm{dist}(x_1) & G(x_1) \\ \vdots & \vdots & \vdots \\ 1 & -10 \ln_{10} \mathrm{dist}(x_N) & G(x_N) \end{bmatrix},$$

$$\boldsymbol{\alpha} = \begin{bmatrix} p_t \\ \kappa \\ \varsigma \end{bmatrix}, \quad t(x) = \begin{bmatrix} 1 \\ -10 \ln_{10} \mathrm{dist}(x) \\ G(x) \end{bmatrix}.$$

We use the mean value (8) as the estimator $\hat{Z}(x)$ for the unknown quantity $Z(x)$. Note that the estimation of $(Z(x_1), \ldots, Z(x_N))^T$ is not $\mathbf{Y}$ since at locations where we have measurements, the prediction technique (8) acts as a denoising algorithm. The prediction formula (8) involves the inversion of the matrix $\boldsymbol{\Sigma}$. By using standard matrix formulas (see e.g. [26, Section 1.5 , Eq. (18)]) we have

$$\boldsymbol{\Sigma}^{-1} = \sigma^{-2}\boldsymbol{I}_N - \sigma^{-2}\boldsymbol{S}\left\{\sigma^2 \boldsymbol{K}^{-1} + \boldsymbol{S}^T\boldsymbol{S}\right\}^{-1} \boldsymbol{S}^T. \quad (10)$$

The key property of this FRK model is that it only requires the inversion of $r \times r$ matrices. Therefore, the computational cost for the REM prediction is $O(r^2\, N)$ which is a drastic reduction when compared to the classical Kriging in situations when $N$ is large. The prediction formula also requires the knowledge of $(\boldsymbol{\alpha}, \sigma^2, \boldsymbol{K})$. The goal of the following section is to address the estimation of these parameters.

### C. Parameter estimation of the Fixed Rank Kriging model

We first expose the method described in the original paper devoted to the FRK model [14]. We also provide a rigorous proof of some weaknesses of this estimation technique pointed out in [27] through numerical experiments. We then propose a second method which is more robust.

*1) Parameter estimation by a method of moments:* In [14], $\boldsymbol{\alpha}$ is estimated by the weighted least squares estimator: given an estimation $(\hat{\sigma}^2, \widehat{\boldsymbol{K}})$ of $(\sigma^2, \boldsymbol{K})$ which yields an estimation $\widehat{\boldsymbol{\Sigma}}$ of $\boldsymbol{\Sigma}$ (see Eq. (7)), we have $\hat{\boldsymbol{\alpha}}_{\mathrm{WLS}} = (\boldsymbol{T}^T \widehat{\boldsymbol{\Sigma}}^{-1} \boldsymbol{T})^{-1} \boldsymbol{T}^T \widehat{\boldsymbol{\Sigma}}^{-1} \mathbf{Y}$. Parameters $\sigma^2$ and $\boldsymbol{K}$ are estimated by a method of moments: the $N$ observations are replaced with $M$ "pseudo-observations" located at $x'_1, \cdots, x'_M$ in $\mathbb{R}^2$. For each $i = 1, \cdots, M$, a pseudo-observation is constructed as the average of the initial observations $Y(x_\ell), \ell = 1, \cdots, N$ which are in a neighborhood of $x'_i$. The parameter $M$ is chosen by the user such that $r < M << N$. An empirical $M \times M$ covariance matrix $\widehat{\boldsymbol{\Sigma}}_M$ is then associated to these pseudo-observations; it is easily invertible due to its reduced dimensions. Finally, the same "binning" technique is applied to the matrix $\boldsymbol{S}$ which yields a $M \times r$ matrix $\boldsymbol{S}_M$ (see [14, Section 3.3.] for a detailed construction of $\widehat{\boldsymbol{\Sigma}}_M$ and $\boldsymbol{S}_M$; see also Appendix A below for a partial description). $\sigma^2, \boldsymbol{K}$ are then estimated by (see [14, Eq. (3.10)] applied with $\bar{V} = \boldsymbol{I}_M$ and $\bar{S} = \boldsymbol{S}_M$)

$$\hat{\sigma}^2 = \frac{\mathrm{Tr}\left(\left(\boldsymbol{I}_M - \boldsymbol{Q}\boldsymbol{Q}^T\right)\widehat{\boldsymbol{\Sigma}}_M\right)}{\mathrm{Tr}\left(\boldsymbol{I}_M - \boldsymbol{Q}\boldsymbol{Q}^T\right)}, \quad (11)$$

$$\widehat{\boldsymbol{K}} = \boldsymbol{R}^{-1}\boldsymbol{Q}^T(\widehat{\boldsymbol{\Sigma}}_M - \hat{\sigma}^2 \boldsymbol{I}_M)\boldsymbol{Q}(\boldsymbol{R}^{-1})^T, \quad (12)$$

where Tr denotes the trace and $\boldsymbol{S}_M = \boldsymbol{Q}\boldsymbol{R}$ is the orthogonal-triangular decomposition of $\boldsymbol{S}_M$ ($\boldsymbol{Q}$ is a $M \times r$ matrix which contains the first $r$ columns of a unitary matrix and $\boldsymbol{R}$ is an invertible upper triangular matrix). These estimators are obtained by fitting $\sigma^2 \boldsymbol{I}_M + \boldsymbol{S}_M \boldsymbol{K}\boldsymbol{S}_M^T$ to $\widehat{\boldsymbol{\Sigma}}_M$, solving the optimization problem $\min_{\sigma^2, \boldsymbol{K}} \|\widehat{\boldsymbol{\Sigma}}_M - \sigma^2 \boldsymbol{I}_M - \boldsymbol{S}_M \boldsymbol{K}\boldsymbol{S}_M^T\|$ where in this equation, $\|\cdot\|$ denotes the Froebenius norm (to have a better intuition of this strategy, compare this criterion to Eq. (7)). $\widehat{\boldsymbol{K}}$ has to be positive definite since it estimates an invertible covariance matrix. In [27], the authors observe through numerical examples that the estimator (12) is a singular covariance matrix (hence, they introduce an "eigen-value lifting" procedure to modify (12) and obtain a positive definite matrix (see [27, Section 3.2.])). We identify sufficient conditions for this empirical observation to be always valid. More precisely, we establish in Appendix A the following,

*Proposition 1:* Assume that $\boldsymbol{S}_M$ is a full rank matrix and let $\boldsymbol{S}_M = \boldsymbol{Q}\boldsymbol{R}$ be its orthogonal-triangular decomposition ($\boldsymbol{Q}$ is a $M \times r$ matrix which collects the first $r$ columns of a unitary matrix). Denote by $(\lambda_j)_j$ the eigenvalues of $\widehat{\boldsymbol{\Sigma}}_M$ and $\mathsf{V}_j$ the eigenspace of $\lambda_j$. Then

(i) $\widehat{\boldsymbol{\Sigma}}_M$ is positive semi-definite.
(ii) $\hat{\sigma}^2$ given by (11) is lower bounded by $\inf_{j:\exists v \in \mathsf{V}_j, \|\boldsymbol{Q}^T v\| < \|v\|} \lambda_j$.
(iii) $\widehat{\boldsymbol{K}}$ given by (12) is positive definite iff $\hat{\sigma}^2 \in [0, \lambda_{\min}(\boldsymbol{Q}^T \widehat{\boldsymbol{\Sigma}}_M \boldsymbol{Q}))$ where $\lambda_{\min}(A)$ denotes the minimal eigenvalue of $A$.

We also give in Appendix A a sufficient condition which implies that the minimal eigenvalue (say $\lambda_1$) of $\widehat{\boldsymbol{\Sigma}}_M$ is positive. If there exists $v \in \mathsf{V}_i$ such that $\|\boldsymbol{Q}^T v\| = \|v\|$ then $\boldsymbol{Q}^T v$ is an eigenvector of $\boldsymbol{Q}^T \widehat{\boldsymbol{\Sigma}}_M \boldsymbol{Q}$ associated to the eigenvalue $\lambda_i$ (observe indeed that if $\|\boldsymbol{Q}^T v\| = \|v\|$, then there exists $\mu \in \mathbb{R}^r$ such that $v = \boldsymbol{Q}\mu$ and this vector satisfies $\mu = \boldsymbol{Q}^T v$). Therefore, if $\lambda_1 > 0$ and for any $v \in \mathsf{V}_1$, $\|\boldsymbol{Q}^T v\| = \|v\|$ then Proposition 1 implies that $\widehat{\boldsymbol{K}}$ given by (12) can not be positive definite.

*2) Parameter estimation by Maximum Likelihood:* We propose to estimate the parameters by the Maximum Likelihood Estimator (MLE), following an idea close to that of [28], [29]. Observe from (5) and (6) that $\mathbf{Y} = \boldsymbol{T}\boldsymbol{\alpha} + \boldsymbol{S}\boldsymbol{\eta} + \sigma\boldsymbol{\varepsilon}$ with $\boldsymbol{\varepsilon} = (\varepsilon(x_1), \cdots, \varepsilon(x_N))^T$. This equation shows that from $\mathbf{Y}$, it is not possible to estimate a general covariance matrix $\boldsymbol{K}$ since roughly speaking, $\mathbf{Y}$ is obtained from a single realization of a Gaussian vector $\boldsymbol{\eta}$ with covariance matrix $\boldsymbol{K}$. Therefore, we introduce a parametric model for this covariance matrix, depending on some vector $\upsilon$ of low dimension: we will write $\boldsymbol{K}(\upsilon)$. We give an example of such a parametric family in Section IV-B2; see also [25, Chapter 4].

Since $\boldsymbol{\eta}$ and $(\varepsilon(x))_x$ are independent processes, $\mathbf{Y}$ is a $\mathbb{R}^N$-valued Gaussian vector with mean $\boldsymbol{T}\boldsymbol{\alpha}$ and with covariance matrix $\boldsymbol{\Sigma} = \sigma^2 \boldsymbol{I}_N + \boldsymbol{S}\boldsymbol{K}(\upsilon)\boldsymbol{S}^T$. Therefore the log-likelihood $L_{\mathbf{Y}}(\boldsymbol{\theta})$ of the observations $\mathbf{Y}$ given the parameters $\boldsymbol{\theta} = (\boldsymbol{\alpha}, \sigma^2, \upsilon)$ is, up to an additive constant,

$$L_{\mathbf{Y}}(\boldsymbol{\theta}) = -\frac{1}{2}\ln\det(\sigma^2 \boldsymbol{I}_N + \boldsymbol{S}\boldsymbol{K}(\upsilon)\boldsymbol{S}^T) - \frac{(\mathbf{Y} - \boldsymbol{T}\boldsymbol{\alpha})^T}{2\sigma^2}\left(\boldsymbol{I}_N - \boldsymbol{S}\left\{\sigma^2 \boldsymbol{K}^{-1}(\upsilon) + \boldsymbol{S}^T\boldsymbol{S}\right\}^{-1}\boldsymbol{S}^T\right)\cdots \times (\mathbf{Y} - \boldsymbol{T}\boldsymbol{\alpha}) \ , \tag{13}$$

where we used (10) for the expression of $\boldsymbol{\Sigma}^{-1}$. Maximizing directly the log-likelihood function $\boldsymbol{\theta} \mapsto L_{\mathbf{Y}}(\boldsymbol{\theta})$ is not straightforward and cannot be computed analytically. We therefore propose a numerical solution based on the Expectation Maximization (EM) algorithm [30]. EM allows the computation of the MLE in latent data models; in our framework, the latent variable is $\boldsymbol{\eta}$. It is an iterative algorithm which produces a sequence $(\boldsymbol{\theta}_{(l)})_{l \geq 0}$ satisfying $L_{\mathbf{Y}}(\boldsymbol{\theta}_{(l+1)}) \geq L_{\mathbf{Y}}(\boldsymbol{\theta}_{(l)})$. This property is fundamental for the proof of convergence of any EM sequence [31]. Each iteration of EM consists in two steps: an Expectation step (E-step) and a Maximization step (M-step). Given the current value $\boldsymbol{\theta}_{(l)}$ of the parameter, the E-step consists in the computation of the expectation of the log-likelihood of $(\mathbf{Y}, \boldsymbol{\eta})$ under the conditional distribution of $\boldsymbol{\eta}$ given $\mathbf{Y}$ for the current value of the parameter $\boldsymbol{\theta}_{(l)}$:

$$Q(\boldsymbol{\theta}; \boldsymbol{\theta}_{(l)}) = \mathbb{E}\left[\ln \Pr(\mathbf{Y}, \boldsymbol{\eta}; \boldsymbol{\theta}) | \mathbf{Y}; \boldsymbol{\theta}_{(l)}\right],$$

where $\theta \mapsto \Pr(\mathbf{Y}, \boldsymbol{\eta}; \boldsymbol{\theta})$ is the likelihood of $(\mathbf{Y}, \boldsymbol{\eta})$. In the M-step, the parameter is updated as the value maximizing $\boldsymbol{\theta} \mapsto Q(\boldsymbol{\theta}; \boldsymbol{\theta}_{(l)})$ or as any value $\boldsymbol{\theta}_{(l+1)}$ satisfying

$$Q(\boldsymbol{\theta}_{(l+1)}; \boldsymbol{\theta}_{(l)}) \geqslant Q(\boldsymbol{\theta}_{(l)}; \boldsymbol{\theta}_{(l)}) \ . \tag{14}$$

The E- and M-steps are repeated until convergence, which in practice may mean when the difference between $\|\boldsymbol{\theta}_{(l)} - \boldsymbol{\theta}_{(l+1)}\|$ changes by an arbitrarily small amount determined by the user (see e.g. [30, Chapter 3]). In our framework, we have

$$\begin{aligned} Q(\boldsymbol{\theta}; \tilde{\boldsymbol{\theta}}) = &-\frac{N}{2}\ln(\sigma^2) - \frac{1}{2}\ln(\det(\boldsymbol{K}(\upsilon))) - \frac{1}{2\sigma^2}\|\mathbf{Y} - \boldsymbol{T}\boldsymbol{\alpha}\|^2 \\ &- \frac{1}{2}\mathrm{Tr}\left(\left(\frac{\boldsymbol{S}^T\boldsymbol{S}}{\sigma^2} + \boldsymbol{K}^{-1}(\upsilon)\right)\mathbb{E}\left[\boldsymbol{\eta}\boldsymbol{\eta}^T | \mathbf{Y}; \tilde{\boldsymbol{\theta}}\right]\right) \\ &+ \frac{1}{\sigma^2}(\mathbf{Y} - \boldsymbol{T}\boldsymbol{\alpha})^T \boldsymbol{S}\mathbb{E}\left[\boldsymbol{\eta} | \mathbf{Y}; \tilde{\boldsymbol{\theta}}\right], \end{aligned} \tag{15}$$

where (see e.g. [12, Appendix C])

$$\begin{aligned} \mathbb{E}\left[\boldsymbol{\eta} | \mathbf{Y}; \tilde{\boldsymbol{\theta}}\right] &= \left(\boldsymbol{S}^T\boldsymbol{S} + \tilde{\sigma}^2 \boldsymbol{K}^{-1}(\tilde{\upsilon})\right)^{-1} \boldsymbol{S}^T (\mathbf{Y} - \boldsymbol{T}\tilde{\boldsymbol{\alpha}}), \\ \mathrm{cov}\left[\boldsymbol{\eta} | \mathbf{Y}; \tilde{\boldsymbol{\theta}}\right] &= \left(\frac{\boldsymbol{S}^T\boldsymbol{S}}{\tilde{\sigma}^2} + \boldsymbol{K}^{-1}(\tilde{\upsilon})\right)^{-1}. \end{aligned}$$

The update formulas of the parameters $(\boldsymbol{\alpha}, \sigma^2)$ are given by (see e.g. [12, Appendix B] for the proof)

$$\begin{aligned} \boldsymbol{\alpha}_{(l+1)} &= \left(\boldsymbol{T}^T\boldsymbol{T}\right)^{-1} \boldsymbol{T}^T \left(\mathbf{Y} - \boldsymbol{S}\,\mathbb{E}\left[\boldsymbol{\eta} | \mathbf{Y}; \boldsymbol{\theta}_{(l)}\right]\right), \\ \sigma^2_{(l+1)} &= \frac{1}{N}\mathbb{E}\left[\left\|\mathbf{Y} - \boldsymbol{T}\boldsymbol{\alpha}_{(l+1)} - \boldsymbol{S}\boldsymbol{\eta}\right\|^2 | \mathbf{Y}; \boldsymbol{\theta}_{(l)}\right]. \end{aligned}$$

With this choice, we have $Q(\boldsymbol{\alpha}_{(l+1)}, \sigma^2_{(l+1)}, \upsilon; \boldsymbol{\theta}_{(l)}) \geq Q(\boldsymbol{\theta}_{(l)}; \boldsymbol{\theta}_{(l)})$, for any $\upsilon$. The update of $\upsilon$ is specific to each parametric model for $\boldsymbol{K}$. Upon noting that the first order derivative of $\upsilon = (\upsilon_1, \cdots, \upsilon_p) \mapsto Q(\boldsymbol{\alpha}, \sigma^2, \upsilon; \boldsymbol{\theta}_{(l)})$ w.r.t. $\upsilon_k$ is given by

$$\begin{aligned} &-\frac{1}{2}\mathrm{Tr}\left(\boldsymbol{K}^{-1}(\upsilon)\frac{\partial \boldsymbol{K}(\upsilon)}{\partial \upsilon_k}\right) \\ &+ \frac{1}{2}\mathrm{Tr}\left(\boldsymbol{K}^{-1}(\upsilon)\mathbb{E}\left[\boldsymbol{\eta}\boldsymbol{\eta}^T | \mathbf{Y}; \boldsymbol{\theta}_{(l)}\right]\boldsymbol{K}^{-1}(\upsilon)\frac{\partial \boldsymbol{K}(\upsilon)}{\partial \upsilon_k}\right), \end{aligned} \tag{16}$$

$\upsilon_{(l+1)}$ can be defined as the unique root of this gradient whenever it is the global maximum. Another strategy is to perform one iteration of a Newton-Raphson algorithm starting from $\upsilon_{(l)}$ with a step size chosen in order to satisfy the EM condition (14). See e.g. [30, Section 4.14] for EM combined with Newton-Raphson procedures. In Section IV-B2, we will give an example of structured covariance matrix and will derive the Newton-Raphson strategy to update one of the parameters.

## III. REM EXTENDED TO MULTICELLULAR NETWORK

We now consider a multicellular LTE network. In real network, UEs measure the received power of several BSs in order to choose the best serving one: the UE, this procedure is called the *cell selection*. In LTE, cell selection is applied by comparing the instant measured RSRP from all potential cells and choosing the cell providing the highest RSRP value [32]. In this section, we adapt the FRK model and the REM prediction technique described in Section II-B in order to address this multicellular use-case.

We assume that the reported measurements correspond to the RSRP of the best serving cell: each measurement consists in the RSRP measure, the location information and the corresponding cell identifier (CID). The received power $Z_i(x)$ from the $i$-th BS at location $x$ is given by $Z_i(x) = 0$ is $x \notin \mathcal{D}_i$ and if $x \in \mathcal{D}_i$,

$$Z_i(x) = p_{t,i} - 10\kappa_i \ln_{10}(\text{dist}_i(x)) + \varsigma_i G_i(x) + \boldsymbol{s}_i(x)^T \boldsymbol{\eta}_i \quad (17)$$

where $\mathcal{D}_i \subseteq \mathbb{R}^2$, $p_{t,i}$ is the transmitted power of the $i$-th BS, $\kappa_i$ is the path loss exponent corresponding to the $i$-th BS and $\text{dist}_i(x)$ is the distance from $x$ to the $i$-th BS. We can choose $\mathcal{D}_i \neq \mathbb{R}^2$ to model geographic area which are not covered by the $i$-th BS. $\boldsymbol{\eta}_i$ is a Gaussian variable with zero mean and covariance matrix $\boldsymbol{K}_i$. $\boldsymbol{s}_i(x) : \mathbb{R}^2 \to \mathbb{R}^{r_i}$ collects $r_i$ deterministic spatial basis functions.

$\varsigma_i G_i(x)$ is the antenna gain which depends on the mobile location $x$. In our use-case, the antennas used for each BS are tri-sectored; we use a typical antenna pattern proposed in the 3GPP standard [1] with a horizontal gain only since we are using a 2-dimensional model:

$$G_i(x) = -\min\left[12\left(\frac{\psi_{x,i}}{\psi_{3dB}}\right)^2, A_m\right], \quad (18)$$

where $\psi_{x,i}$ is the angle between the UE location $x$, and the $i$-th BS antenna azimuth. $\psi_{3dB}$ denotes the angle at which the antenna efficiency is $50\%$ and $A_m$ is the maximum antenna gain. For a tri-sectorial antenna, the parameter $\psi_{3dB}$ is usually taken equal to $65°$ and $A_m = 30$dB.

We have $N_i$ observations $Y_i(x)$ having the $i$-th BS as the best serving cell. They are located at $x_{1,i}, \cdots, x_{N_i,i}$ and are noisy measurements of $Z_i(x)$: $Y_i(x) = Z_i(x) + \sigma_i \varepsilon_i(x)$ where $(\varepsilon_i(x))_x$ is a zero mean standard Gaussian process, independent of $\boldsymbol{\eta}_i$. Following the same lines as in section II-B, we define the $N_i \times 1$ column vector $\mathbf{Y}_i = (Y_i(x_{1,i}), \cdots, Y_i(x_{N_i,i}))^T$, and have $\mathbf{Y}_i = \boldsymbol{T}_i\boldsymbol{\alpha}_i + \boldsymbol{S}_i\boldsymbol{\eta}_i + \sigma_i\boldsymbol{\varepsilon}_i$ where

$$\boldsymbol{T}_i = \begin{bmatrix} 1 & -10\ln_{10}(\text{dist}_i(x_{1,i})) & G_i(x_{1,i}) \\ \vdots & \vdots & \vdots \\ 1 & -10\ln_{10}(\text{dist}_i(x_{N_i,i})) & G_i(x_{N_i,i}) \end{bmatrix},$$

$$\boldsymbol{\alpha}_i = \begin{bmatrix} p_{t,i} \\ \kappa_i \\ \varsigma_i \end{bmatrix}, \qquad \boldsymbol{\varepsilon}_i = \begin{bmatrix} \varepsilon_i(x_{1,i}) \\ \vdots \\ \varepsilon_i(x_{N_i,i}) \end{bmatrix}.$$

The parameters $p_{t,i}, \kappa_i, \sigma_i, \varsigma_i$ and $\boldsymbol{K}_i$ are unknown and are estimated from $\mathbf{Y}_i$ by applying the EM technique described in Section II-C (see also Section IV-B for the implementation).

For any $x$ such that $x \in \mathcal{D}_i$, set $\hat{Z}_i(x) = \mathbb{E}\left[Z_i(x)|\mathbf{Y}_i\right]$, the expression of which can easily be adapted from (8). In the multicellular case, the inter-site shadowing correlation can be explained by a partial overlap of the large-scale propagation medium as explained in [33]. Hence, for any $x$ such that $x \in \mathcal{D}_i$, we write $Z_i(x) = Z_i'(x) + W(x)$, where $W(x)$ is the random cross-correlated shadowing term which depends only on the mobile location (also called overlapping propagation term) and $Z_i'(x)$ is the random correlated shadowing related to the $i$-th BS at the location $x$ (also called non-overlapping propagation term). As explained in [33], the r.v. $(Z_i'(x))_i$ are independent, which implies that the probability that a UE located at $x$ is attached to the $i$-th BS (which is denoted by $\text{CID}(x) = i$) is given by

$$\mathbb{P}(\text{CID}(x) = i) = \mathbb{E}\left[\prod_{j \neq i : x \in \mathcal{D}_j} \mathbb{1}_{Z_j(x) \leq Z_i(x)}\right]. \quad (19)$$

A simple approximation consists in approximating this expectation by

$$\prod_{j \neq i : x \in \mathcal{D}_j} \mathbb{1}_{\hat{Z}_j(x) \leq \hat{Z}_i(x)}.$$

This yields the estimation rules for the CID and the RSRP value at $x$

$$\widehat{\text{CID}}(x) = \text{argmax}_{j : x \in \mathcal{D}_j} \hat{Z}_j(x),$$
$$\hat{Z}(x) = \hat{Z}_{\widehat{\text{CID}}(x)}(x) = \max_{j : x \in \mathcal{D}_j} \hat{Z}_j(x).$$

## IV. APPLICATIONS TO CELLULAR COVERAGE MAP

### A. Data sets description

For the single cell use-case, we consider a simulated data set and a real data set. The first data set consists of simulated measurement points generated with a very accurate planning tool, which uses a sophisticated ray-tracing propagation model developed for operational network planning [34]. This data is considered as the ground-truth of the coverage in the area of interest. The collected data set corresponds to the LTE RSRP values in an urban scenario located in the Southwest of Paris (France). The environment is covered by a macro-cell with an omni-directional antenna. These measurement points are located on a $1000$ m$\times$$1000$ m surface, regularly spaced on a cartesian grid consisting of $5$ m $\times 5$ m squares; this yields a total of $40401$ measurement points (see Fig. 1a, where the antenna location is $(595\,416\,\text{m}, 2\,425\,341\,\text{m})$). In order to model the noise measurements, a zero mean Gaussian noise with variance equal to 3 dB is added to the simulated measurements. This yields what we called in Section II the process $\{Y(x), x \in \mathcal{D}\}$, where $\mathcal{D} \subset \mathbb{R}^2$.

The second data set corresponds to real measurement points reported from Drive Tests (DT) done by Orange France teams, in a rural area located in southwestern France. The BS is about 30 m height and covers an area of 22 km$\times$10 km. 7800 measurement points have been collected in the 800

MHz frequency band using a typical user's mobile device connected to a software tool for data acquisition.The locations of the measurement points are shown on Fig. 1b - note that they are along the roads and the antenna is located at $(408\,238\,\text{m}, 1\,864\,600\,\text{m})$. For the multicellular use-case, we consider a simulated data set provided by the aforementioned Orange planning tool. This planning tool calculates RSRP values in a sub-urban environment shown in Fig. 2a, consisting of 12 antennas grouped into 4 sites of 3 directional antennas. The inter-site distance is bigger than 1 km. The antennas are tri-sectored. The RSRP values are computed over a regular grid of size 25 m×25 m over a 12.4 km$^2$ geographic area, which results in a total of 20 008 locations; and it is realized over a 2.6 GHz frequency band. The planning tool returns, at each location of the regular grid, both the RSRP value and the ID of the best serving cell. Fig. 2b displays the RSRP values and Fig. 2c shows the best serving cell map where each color corresponds to a cell coverage area.

## B. EM implementation

*1) Choice of the basis functions $\boldsymbol{s}$:* The basis functions $x \mapsto \boldsymbol{s}(x) = (S_1(x), \ldots, S_r(x))$ and their number $r$ both control the complexity and the accuracy of the FRK prediction technique. Following the suggestions in [14], we choose the $l$-th basis function $x \mapsto S_l(x)$ as a symmetric function centered at locations $x'_l$: $S_l$ is a bi-square function defined as

$$S_l(x) = \begin{cases} \left[1 - (\|x - x'_l\| / \tau)^2\right]^2, & \text{if } \|x - x'_l\| \leqslant \tau \ , \\ 0, & \text{otherwise} \ . \end{cases} \tag{20}$$

The parameter $\tau$ controls the support of the function. In the numerical applications below, the centers of the basis functions $x'_l$ and their number $r$ are chosen as follows: $r_{\max}$ functions are located on a Cartesian grid where the elements are $\tau \times \tau$ squares covering the whole geographic area of interest. Then, for each function $S_l$, if none of the $N$ locations $x_1, \cdots, x_N$ is in a $\tau$-neighborhood of the center $x'_l$, this function is removed. The number of the remaining basis function is $r$. On Fig. 3a and Fig. 3b, we show the locations of the $N$ observations (red circle) and the locations of the $r$ basis function centers (blue crosses) for two different data sets. In Fig. 3a, $\tau = 100$ m and $r = r_{\max}$ (and $N = 2000$) while in Fig. 3b, $\tau = 250$ m, $r_{\max} = 2660$ and $r = 467$.

*2) Structured covariance matrix $\boldsymbol{K}$:* Several examples of structured covariance matrix $\boldsymbol{K}$ can be chosen. In the radio cellular context, the shadowing term can be modeled as a zero-mean Gaussian random variable with an exponential correlation model [35]. Thus, $\boldsymbol{K}$ is given by

$$\boldsymbol{K}(\beta, \phi) = \frac{\tilde{\boldsymbol{K}}(\phi)}{\beta}, \tag{21}$$

$$\text{with } \tilde{K}_{i,j}(\phi) = \exp\left(-\frac{\|x'_i - x'_j\|}{\exp(\phi)}\right), \tag{22}$$

where $\|x'_i - x'_j\|$ is the Euclidean distance between the two locations $x'_i$ and $x'_j$ (related to the basis functions, see Section IV-B1). $1/\beta$ and $\exp(\phi)$ are respectively the variance of $\eta_l$, $1 \le l \le r$; and a rate of decay of the correlation (the choice of the parametrization $\exp(\phi)$ avoids the introduction of a constraint of sign when estimating $\phi$). We therefore have $\upsilon = (\beta, \phi) \in \mathbb{R}_\star^+ \times \mathbb{R}$. For this specific parametric matrix (21-22), a possible update of the parameters $(\beta, \phi)$ which ensures the monotonicity property of the EM algorithm is (see e.g. [12, Appendix B]): $\beta_{(l+1)} = r/\mathrm{Tr}\left(\tilde{\boldsymbol{K}}_{(l)}^{-1}\ \mathcal{V}_{(l)}\right)$ and

$$\phi_{(l+1)} = \phi_{(l)} - \frac{a_{(l)}}{\mathcal{H}_{(l)}} \cdots \\ \times \mathrm{Tr}\left(\left(\beta_{(l+1)}\tilde{\boldsymbol{K}}_{(l)}^{-1}\mathcal{V}_{(l)} - \boldsymbol{I}_r\right)\tilde{\boldsymbol{K}}_{(l)}^{-1}\ \Delta \circ \tilde{\boldsymbol{K}}_{(l)}\right)$$

where $\tilde{\boldsymbol{K}}_{(l)}$ is a shorthand notation for $\tilde{\boldsymbol{K}}(\phi_{(l)})$, $\Delta$ is the $r \times r$ matrix with entries $(\|x'_i - x'_j\|)_{ij}$, $\mathcal{V}_{(l)}$ is a shorthand notation for $\mathbb{E}\left[\boldsymbol{\eta}\boldsymbol{\eta}^T|\mathbf{Y};\boldsymbol{\theta}_{(l)}\right]$, $\circ$ denotes the Hadamard product and

$$\mathcal{H}_{(l)} = -\mathrm{Tr}\left(\tilde{\boldsymbol{K}}_l^{-1}\ \Delta \circ \tilde{\boldsymbol{K}}\left(\beta_{(l+1)}\tilde{\boldsymbol{K}}_l\mathcal{V}_{(l)} - \boldsymbol{I}_r\right)\right) \\ + \exp(-\phi_{(l)})\mathrm{Tr}\left(\tilde{\boldsymbol{K}}_l^{-1}\ \Delta \circ \Delta \circ \tilde{\boldsymbol{K}}_l\left(\beta_{(l+1)}\tilde{\boldsymbol{K}}_l\mathcal{V}_{(l)} - \boldsymbol{I}_r\right)\right) \\ + \exp(-\phi_{(l)})\mathrm{Tr}\left(\left(\tilde{\boldsymbol{K}}_l^{-1}\ \Delta \circ \tilde{\boldsymbol{K}}_l\right)^2\left(\boldsymbol{I}_r - 2\beta_{(l+1)}\tilde{\boldsymbol{K}}_l\mathcal{V}_{(l)}\right)\right);$$

$a_{(l)} \in (0,1)$ is chosen so that $Q(\boldsymbol{\theta}_{(l+1)};\boldsymbol{\theta}_{(l)}) \geq Q(\boldsymbol{\theta}_{(l)};\boldsymbol{\theta}_{(l)})$.

*3) EM convergence:* EM converges whatever the initial value $\boldsymbol{\theta}_{(0)}$ (see [31]); the limiting points of the EM sequences are the stationary points of the log-likelihood of the observations $\mathbf{Y}$. We did not observe that the initialization $\boldsymbol{\theta}_{(0)}$ plays a role on the limiting value of our EM runs. A natural initial value for $\boldsymbol{\alpha}$ is the Ordinary Least Square estimator given by $\boldsymbol{\alpha}_{(0)} = \left(\boldsymbol{T}^T\boldsymbol{T}\right)^{-1}\boldsymbol{T}^T\mathbf{Y}$. We choose $\phi_{(0)}$ large enough so that the matrix $\tilde{\boldsymbol{K}}(\phi_{(0)})$ looks like the identity matrix; in practice, we choose $\tau/\exp(\phi)$ in the order of 5. Finally, we compute the empirical variance $\mathcal{V}$ of the components of the residual vector $\mathbf{Y} - \boldsymbol{T}\boldsymbol{\alpha}_{(0)}$ and choose $\beta_{(0)}^{-1} + \sigma_{(0)}^2 = \mathcal{V}$; roughly speaking, we start from a model with uncorrelated shadowing term. The algorithm is stopped when $\left\|\boldsymbol{\theta}_{(l)} - \boldsymbol{\theta}_{(l-1)}\right\| < 10^{-5}$ over 100 successive iterations. We report in Table I the values of the parameters at convergence of EM for the simulated data set.

TABLE I. SIMULATED DATA SET, WHEN $\tau = 50$ M, $r = 400$ AND $N = 32000$

| $\hat{\sigma}^2$ | $\hat{\boldsymbol{\alpha}}$ | | $1/\hat{\beta}$ | $\hat{\phi}$ |
|---|---|---|---|---|
| 18.15 | −49.55 | 2.73 | 12.5 | 3.63 |

## C. Prediction Error Analysis for the single cell use-case

Each data set is splitted into a learning set and a test set. Using the data in the learning set, the parameters are estimated by the method described in Section II-C. The performances are then evaluated using the data in the test set. In order to make this analysis more robust to the choice of the learning and test sets, we perform a $k$-fold cross validation [36] (here, we choose $k = 5$) with a uniform data sampling of the

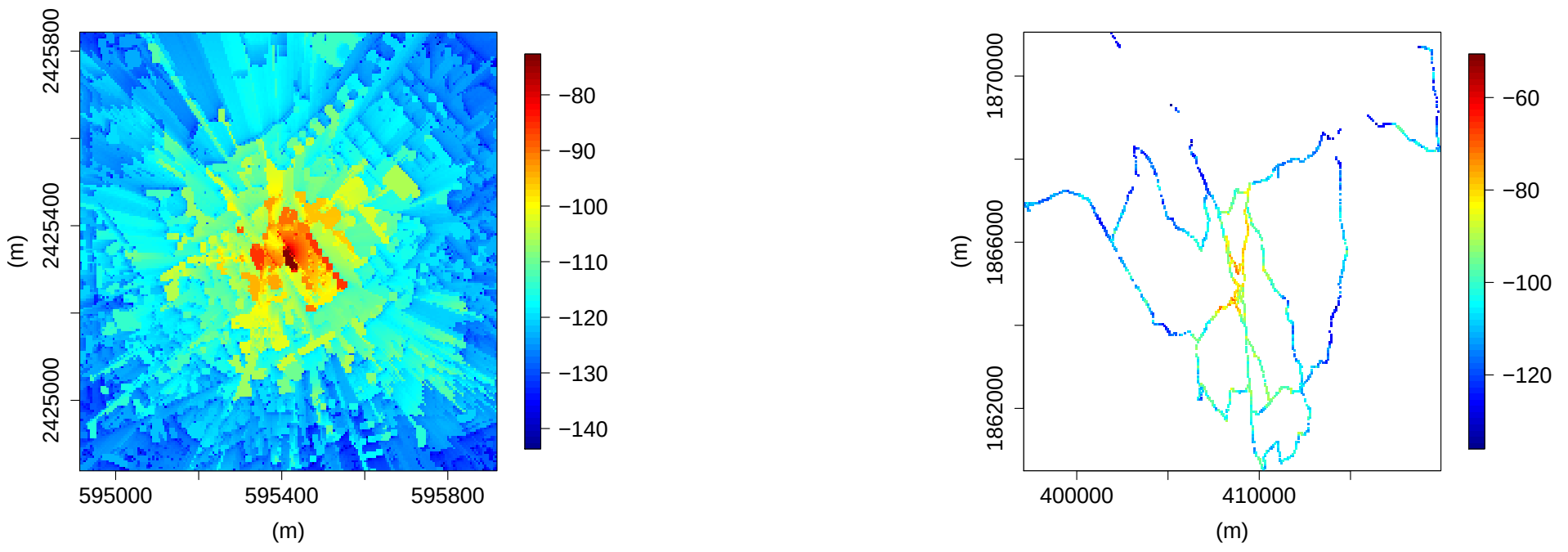


(a) Simulated data set.

(b) Rural data set.

Fig. 1. One cell case: the measurements $(Y(x))_x$.

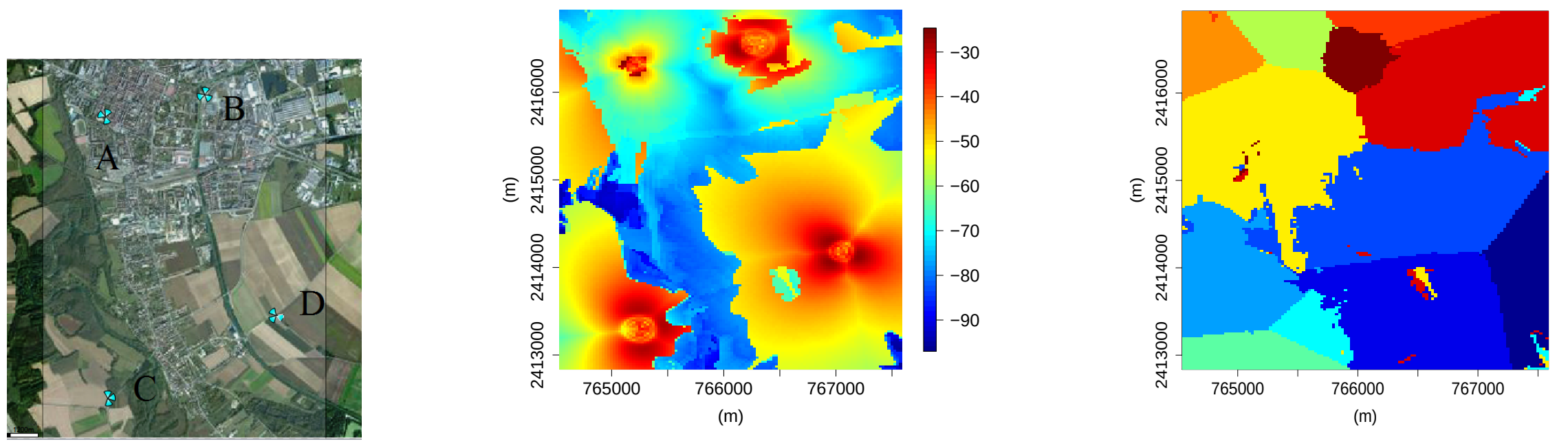


(a)

(b)

(c)

Fig. 2. Multicellular case: (a) BS locations; (b) the simulated RSRP map; (c) measurements grouped in 12 clusters, according to their best serving cell ID

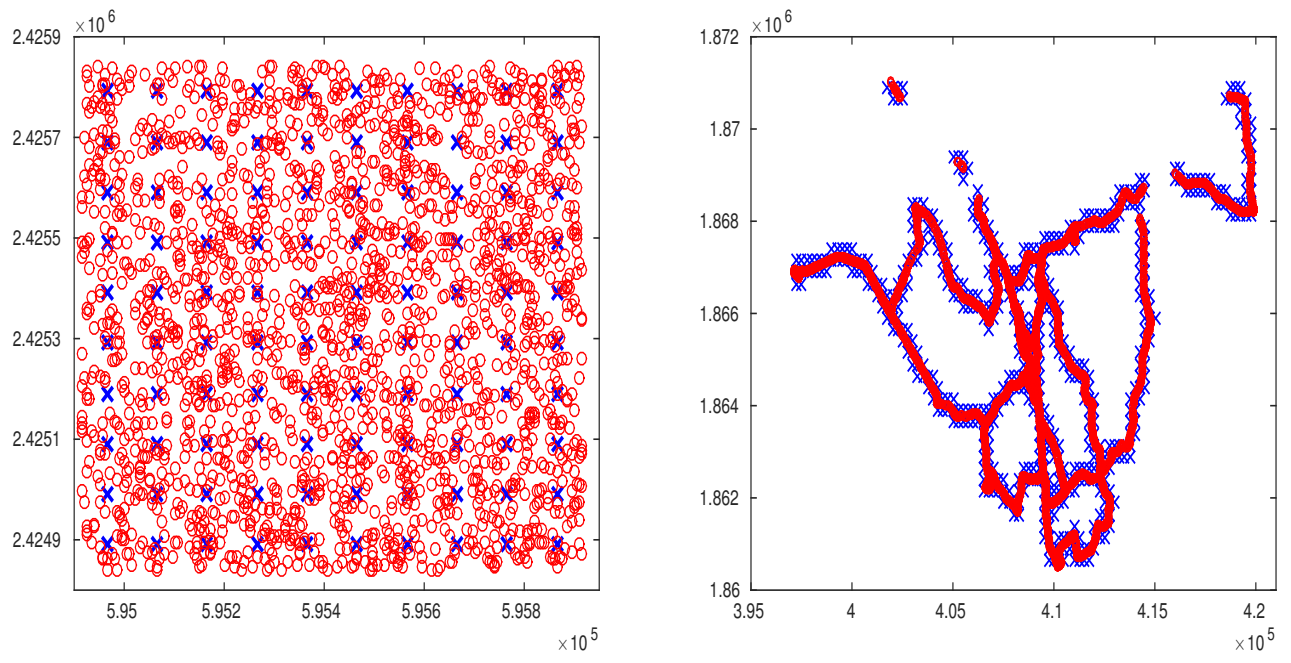

(a) Simulated data set

(b) Real data set

Fig. 3. Locations of the $N$ observations (red circles) and locations of the $r$ centers $x'_l$ (blue crosses) of the basis functions.

subsets (typical values for $k$ are in the range 3 to 10 [25, see Section 5.3.]). Therefore, at each step of this cross-validation procedure, we have a learning set consisting of 80% of the available measurement points (making a learning sets with resp. 32000 and 6000 points for resp. the simulated data set and the real data set).

In order to evaluate the prediction accuracy, we compare the measurements $Y(x)$ to the predicted values $\hat{Y}(x)$ from the model (6). We consider the locations $x$ in the test set $\mathcal{T}$. The model (6) implies that the conditional expectation of $Y(x)$ given $\mathbf{Y}$ at such locations $x$ is equal to the conditional expectation of $Z(x)$ given $\mathbf{Y}$ since $\varepsilon(x)$ is independent of $\mathbf{Y}$. Therefore, for any $x \in \mathcal{T}$, the error (with sign) is $\hat{Y}(x) - Y(x) = \hat{Z}(x) - Y(x)$ where $\hat{Z}(x)$ is given by (8). We evaluate the Root Mean Square Error (RMSE) which is a commonly used prediction error indicator (see e.g. [37]), defined as

$$\text{RMSE} = \left[ \frac{1}{|\mathcal{T}|} \sum_{x \in \mathcal{T}} \left( \hat{Y}(x) - Y(x) \right)^2 \right]^{\frac{1}{2}}, \tag{23}$$

where $|\mathcal{T}|$ denotes the number of observations in the test set $\mathcal{T}$. The RMSE is computed for each of the $k$ successive test sets in the cross-validation analysis. In Tables II and III, we report the mean value of the RMSE over the $k$ partitions and its standard deviation in parenthesis. We compare different strategies for modeling the observations $(Y(x))_x$, for the

parameter estimation of the model and for the prediction:

- `Log-Normal`: the log-normal shadowing model (see (2)) when the parameters $p_t, \kappa, \sigma^2$ are estimated by MLE. $\hat{Z}(x)$ is given by $\hat{p}_t - 10\hat{\kappa}\log_{10}(\text{dist}(x))$; this method does not depend on $r$.
- `FRK`: the FRK model (see section II-B) when the parameters are estimated by MLE (see Sections II-C and IV-B) and $\hat{Z}(x)$ is given by (8), for different values of $r$.

In tables II and III, we report the mean RMSE over the $k$ splits of the data set and its standard deviation between parenthesis. These tables show that the FRK model improves on the log-normal model. For the real data set, it yields a considerably low RMSE (in the order of $3-5$ dB) when compared to the log-normal shadowing model which has a RMSE in the order of 9 dB. For the simulated data set, we have a similar behavior.

TABLE II. SIMULATED DATA SET: MEAN RMSE AND STANDARD DEVIATION IN PARENTHESIS.

| Log-Normal | FRK $r = 1089$ | FRK $r = 100$ |
|---|---|---|
| 5.08 (6.08e-02) | 3.98 (5.18e-02) | 4.67 (4.46e-02) |

TABLE III. REAL DATA SET: MEAN RMSE AND STANDARD DEVIATION IN PARENTHESIS

| Log-Normal | FRK $r = 1000$ | FRK $r = 150$ |
|---|---|---|
| 8.95 (1.46e-01) | 3.51 (1.24e-01) | 5.57 (6.23e-02) |

The computational complexity of the FRK approach is essentially related to $r$, the number of basis functions. On the one hand, the computational cost increases with $r$ and on the other hand, the prediction accuracy increases with $r$. We report on Fig. 4 the running time and the prediction accuracy measured in terms of mean RMSE over the $k$ splitting of the data set into a learning and a test set, as a function of $r$; by convention, the running time is set to 1 when $r = 64$. The plot is obtained with 7 different analysis, obtained with $\tau \in \{30, 40, 50, 60, 80, 100, 120\}$ - or equivalently, $r \in \{1089, 625, 400, 289, 169, 100, 64\}$. It shows that the running time is multiplied by a factor 130 and the prediction accuracy is increased by $20\%$ when moving from $\tau = 120$ ($r = 64$) to $\tau = 30$ ($r = 1089$).

### D. Prediction Error Analysis for the multicellular use-case

The data set is splitted into a learning set with 16 000 points and a test set. Based on their best serving cell ID, these 16 000 points are clustered into 12 subsets. The size of these subsets varies between 1000 and 3500. In Fig. 5a a learning subset associated to a given BS is displayed: note that the observations with a given best serving cell ID are not uniformly distributed over the geographical area of interest. We choose the same initial basis functions for the 12 sub-models (defined by Eq.(20) with $\tau = 150$, which yields $r_{\max} = 588$). For each sub-model, some of the basis functions are canceled as described in Section IV-B1 (see the blue circles and black dots in Fig. 5a). Fig. 5b shows the path-loss function $x \mapsto \hat{p}_{t,i} - 10\hat{\kappa}_i \ln_{10} \text{dist}_i(x) + \hat{\varsigma}_i G_i(x)$: note that, as expected, $\boldsymbol{T}_i\boldsymbol{\alpha}_i$ is bigger in the direction of the antenna spread. In Fig. 5c, we display $\{\hat{Z}_i(x), x \in \mathcal{D}_i\}$. $\mathcal{D}_i$ is defined as the area covering the main direction of the $i$-th antenna radiation.

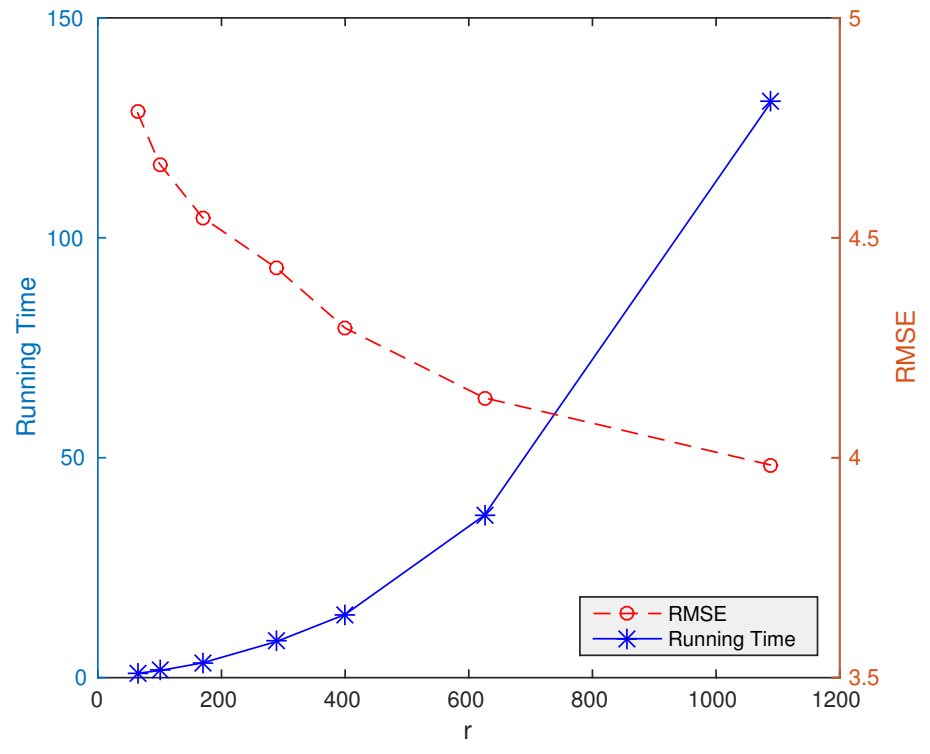


Fig. 4. Simulated Data set: for different values of $r$, the running time and the mean RMSE

The best serving cell ID ($\text{CID}_{bs}$) for any location $x \in \mathcal{D}_i$ is defined as the ID of the BS having the biggest probability that the ME is attached to it at location $x$ as detailed in Eq. 19. Then the predicted received power at location $x$ corresponds to the predicted received power of the best serving cell at that location. For performance evaluation, we first consider an omni-directional antenna model (similar to the one in section IV-C). We compare the predicted cell ID for each location $x$ (that is the index $j$ such that $\hat{Z}(x) = \hat{Z}_j(x)$) to the real one. We obtain an error rate of $53\,\%$ over the locations $x$ in the test set. When we consider the domain clustering introduced in (17) (the antennas are still assumed to be omnidirectional), the error rate on cell ID selection is $31.23\%$ over the test set locations. Finally, we consider the directional model together with the same domain restriction $\mathcal{D}_i$. The error rate is drastically decreased to $12.64\%$. This error rate is expected to further decrease when using real antenna patterns (the impact of approximating real antenna patterns with the 3GPP model is studied for example in [38]).

## V. CONCLUSION

In this work we have studied the performance of the FRK algorithm applied to coverage analysis in cellular networks. This method has a good potential when performing prediction using massive data sets (order of thousands and higher) as it offers a good trade-off between prediction quality and computational complexity compared to classical Kriging techniques. This study has been performed using field-like measurements obtained from an accurate planning tool and real field measurements obtained from drive tests. In addition we have adapted the model to a more practical application: we

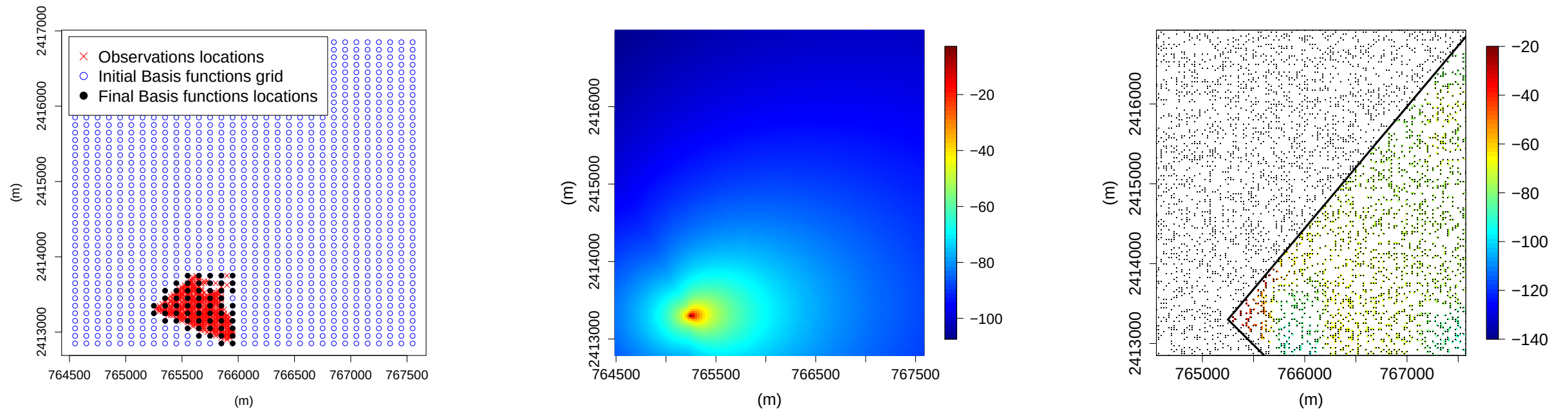


(a) For a given BS: the associated observations $\mathbf{Y}_i$ (red crosses) and the locations of the $r_i$ basis functions (black dots).

(b) $x \mapsto \boldsymbol{T}_i \hat{\boldsymbol{\alpha}}_i$

(c) estimated field $(\hat{Z}_i(x))_x$, for $x \in \mathcal{D}_i$ (the black points correspond to locations $x$ not in $\mathcal{D}_i$).

Fig. 5. Multicellular case: results for a given best serving cell ID $i$

used field-like measurements over several cells with directive antennas. Simulation results show a good performance in terms of coverage prediction and detection of the best serving cell. In future works, we target to further improve this performance by using real antenna patterns. Finally, our ongoing research focuses on extending the model to take into account the location uncertainty and on studying its impact on the prediction performances.

## Appendix A
## Proof of Proposition 1

We recall some notations introduced in [14, Appendix A], which will be useful for the proof of Proposition 1. For $j = 1, \cdots, M$, set $\boldsymbol{W}_j = (W_{j1}, \ldots, W_{jN})^T$, where $W_{lj}$ is the weight associated to the observation $Y(x_j)$ in the neighborhood of the bin center $x'_l$ (see [14] for the expression of these non negative weights). Define the vector of residual $\boldsymbol{D} = (D_1, \cdots, D_N)^T = \mathbf{Y} - \boldsymbol{T}(\boldsymbol{T}^T\boldsymbol{T})^{-1}\boldsymbol{T}^T\mathbf{Y}$, and associate an aggregated vector of residuals $\overline{\boldsymbol{D}} = (\overline{D}_1, \cdots, \overline{D}_M)^T$ and a weighted square residuals

$$\overline{D}_\ell = \frac{\sum_{i=1}^N W_{\ell i} D_i}{\sum_{i=1}^N W_{\ell i}} = \frac{\boldsymbol{W}_\ell^T \boldsymbol{D}}{\boldsymbol{W}_\ell^T 1_N}, \qquad V_\ell = \frac{\sum_{i=1}^N W_{\ell i} D_i^2}{\boldsymbol{W}_\ell^T 1_N}.$$

$1_N$ is the $N \times 1$ vector of ones. The $M \times M$ matrix $\widehat{\boldsymbol{\Sigma}}_M$ is defined by (see [14, Eq. (A.2)])

$$\widehat{\boldsymbol{\Sigma}}_M(l,k) = \overline{D}_\ell \overline{D}_k, \text{ for } l \neq k, \qquad \widehat{\boldsymbol{\Sigma}}_M(k,k) = V_k. \quad (24)$$

*Proof of Proposition 1 (i)* Let $\mu = (\mu_1, \cdots, \mu_M) \in \mathbb{R}^M$. From (24),

$$\mu^T \widehat{\boldsymbol{\Sigma}}_M \mu = \left(\sum_{l=1}^M \mu_l \overline{D}_l\right)^2 + \sum_{l=1}^M \mu_l^2 \left(V_l - \overline{D}_l^2\right) \geq \sum_{l=1}^M \mu_l^2 \left(V_l - \overline{D}_l^2\right).$$

The Jensen's inequality implies that $V_l \geq \overline{D}_l^2$ for any $l$ thus showing that $\mu^T \widehat{\boldsymbol{\Sigma}}_M \mu \geq 0$. Note also that this term is positive for any non null vector $\mu$ iff $V_l - \overline{D}_l^2 > 0$ for any $l$. *(ii)* Since $\widehat{\boldsymbol{\Sigma}}_M$ is a covariance matrix, there exists an orthogonal $M \times M$ matrix $\boldsymbol{U}$ and a diagonal $M \times M$ matrix $\boldsymbol{\Lambda}$ with diagonal entries $(\lambda_i)_i$ such that $\widehat{\boldsymbol{\Sigma}}_M = \boldsymbol{U}\boldsymbol{\Lambda}\boldsymbol{U}^T$. Since $\mathrm{Tr}(AB) = \mathrm{Tr}(BA)$, we have

$$\mathrm{Tr}\left((\boldsymbol{I}_M - \boldsymbol{Q}\boldsymbol{Q}^T)\boldsymbol{U}\boldsymbol{\Lambda}\boldsymbol{U}^T\right) = \sum_{i=1}^M \boldsymbol{B}_{ii}\lambda_i,$$

where $\boldsymbol{B} = \boldsymbol{U}^T(\boldsymbol{I}_M - \boldsymbol{Q}\boldsymbol{Q}^T)\boldsymbol{U}$. Assume that $\boldsymbol{B}_{ii} \geq 0$ for any $i$. Then

$$\mathrm{Tr}\left((\boldsymbol{I}_M - \boldsymbol{Q}\boldsymbol{Q}^T)\boldsymbol{U}\boldsymbol{\Lambda}\boldsymbol{U}^T\right) \geq \left(\inf_{j:\boldsymbol{B}_{jj}>0} \lambda_j\right) \mathrm{Tr}(\boldsymbol{B}).$$

Since $\mathrm{Tr}(\boldsymbol{B}) = \mathrm{Tr}((\boldsymbol{I}_M - \boldsymbol{Q}\boldsymbol{Q}^T)\boldsymbol{U}\boldsymbol{U}^T) = \mathrm{Tr}(\boldsymbol{I}_M - \boldsymbol{Q}\boldsymbol{Q}^T)$, we have $\hat{\sigma}^2 \geq \left(\inf_{j:\boldsymbol{B}_{jj}>0} \lambda_j\right)$. Let us prove that $\boldsymbol{B}_{ii} \geq 0$ for any $i$: for $\mu \in \mathbb{R}^M$, $\mu^T\boldsymbol{B}\mu = \|\boldsymbol{U}\mu\|^2 - \|\boldsymbol{Q}^T(\boldsymbol{U}\mu)\|^2$ and this term is non negative since $\boldsymbol{Q}^T(\boldsymbol{U}\mu)$ is the orthogonal projection of $(\boldsymbol{U}\mu)$ on the column space of $\boldsymbol{Q}$ (or equivalently, of $\boldsymbol{S}_M$). This equality also shows that

$$\{j : \boldsymbol{B}_{jj} > 0\} = \{j : \exists v \in \mathsf{V}_j, \|v\|^2 > \|\boldsymbol{Q}^T v\|^2\} = \{j : \exists v \in \mathsf{V}_j, \|(\boldsymbol{S}_M^\perp)^T v\| > 0\}.$$

(iii) Since $\boldsymbol{S}_M$ is a full rank matrix, $\boldsymbol{R}$ is invertible. Therefore, from (12), it is trivial that $\widehat{\boldsymbol{K}}$ is positive definite iff $\boldsymbol{Q}^T(\widehat{\boldsymbol{\Sigma}}_M - \hat{\sigma}^2\boldsymbol{I}_M)\boldsymbol{Q}$ is positive definite. Since $\boldsymbol{Q}^T\boldsymbol{Q} = \boldsymbol{I}_r$, we have for any $\mu \in \mathbb{R}^r$, $\mu \neq 0$: $\mu^T(\boldsymbol{Q}^T\widehat{\boldsymbol{\Sigma}}_M\boldsymbol{Q} - \hat{\sigma}^2\boldsymbol{I}_r)\mu > 0$ iff $\mu^T(\boldsymbol{Q}^T\widehat{\boldsymbol{\Sigma}}_M\boldsymbol{Q})\mu > \hat{\sigma}^2 \|\mu\|^2$.

*Remark.:* It can be seen from the proof of *(i)* that $\widehat{\boldsymbol{\Sigma}}_M$ is positive definite iff for any $l$, $\boldsymbol{W}_l$ has at least two non null components (say $i_l, j_l$) such that $D_{i_l} \neq D_{j_l}$.

## Acknowledgment

The authors would like to acknowledge Emmanuel De Wailly and Jean-Francois Morlier for their help in data acquisition.